\documentclass[fleqn,usenatbib]{mnras}

\usepackage{newtxtext,newtxmath}
\usepackage[T1]{fontenc}

\DeclareRobustCommand{\VAN}[3]{#2}
\let\VANthebibliography\thebibliography
\def\thebibliography{\DeclareRobustCommand{\VAN}[3]{##3}\VANthebibliography}

\usepackage{graphicx}	% Including figure files
\usepackage{amsmath}	% Advanced maths commands
\usepackage{siunitx}    % units for physical quantities
\usepackage{hyperref}   % links to github
\usepackage{minted}
\usepackage{multirow, makecell}   % used to place images in tensions table
\usepackage[table]{xcolor}     % coloured rows in table
\usepackage{soul}
\DeclareSIUnit\parsec{pc}
\newcommand{\pc}{\textsc{PolyChord}}

\newcommand{\ac}{\textsc{anesthetic}}
\newcommand{\camb}{\textsc{camb}}
\newcommand{\cobaya}{\textsc{Cobaya}}
\newcommand{\lcdm}{$\Lambda$CDM}

\title[Tilting the scales]{Tilting the scales: weighing prior dependency and global tensions of CMB lensing}

\author[A.N.~Ormondroyd et al.]{
A.N.~Ormondroyd,$^{1,2}$\thanks{E-mail: ano23@cam.ac.uk}
W.J.~Handley,$^{1,2}$
M.P.~Hobson$^{1}$
and A.N.~Lasenby$^{1,2}$
\\
$^{1}$Astrophysics Group, Cavendish Laboratory, J.J.~ Thomson Avenue, Cambridge, CB3 0HE, UK\\
$^{2}$Kavli Institute for Cosmology, Madingley Road, Cambridge, CB3 0HA, UK\\
}

\date{Accepted XXX. Received YYY; in original form ZZZ}

\pubyear{2023}

\begin{document}
\label{firstpage}
\pagerange{\pageref{firstpage}--\pageref{lastpage}}
\maketitle

\begin{abstract}
    We provide a nested sampling analysis of the combination of CMB lensing experiments with other cosmological measurements.
    Nested samples can be used to compute global consistency statistics between datasets.
    This is demonstrated for CMB lensing and Baryon Acoustic Oscillations which are uncorrelated, and the correlated case between ACT DR6 and \texttt{NPIPE} lensing.
    We investigate the effect of the prior widths of the spectral tilt $n_\mathrm s$ used in CMB lensing analyses, which quantitatively, but not qualitatively, affect headline constraints.
    In the absence of informative priors, SPT-3G performs more similarly to ACT and \texttt{NPIPE}.
    Bayes factors and the suspiciousness statistic are used to quantify the possibility of tension, and we find the Gaussian assumptions inherent in calculating the suspiciousness tension probability to be unsuitable in the case of strong agreement between CMB lensing experiments.
\end{abstract}

\begin{keywords}
methods: statistical -- cosmological parameters -- gravitational lensing: weak
\end{keywords}

%%%%%%%%%%%%%%%%%%%%%%%%%%%%%%%%%%%%%%%%%%%%%%%%%%

%%%%%%%%%%%%%%%%% BODY OF PAPER %%%%%%%%%%%%%%%%%%

\section{Introduction}

Cosmological datasets are used to constrain the values of parameters of models, such as \lcdm{}.
Individual datasets may only constrain some of the parameters, or have degeneracy along a particular direction in parameter space.
By combining datasets with constraining power along different directions, parameters can be more tightly constrained, breaking the degeneracy.
However, if two datasets are in tension with one another, they will produce a ``suspiciously'' tight constraint when in fact one should conclude that the datasets disagree.
Therefore, care must be taken when combining measurements.

Nested sampling \citep{skilling2004} provides an alternative approach to Metropolis--Hastings methods for producing posterior samples as a by-product of computing the Bayesian evidence.
Nested sampling has no issues with unconstrained parameters, the posterior samples will correspond to their prior without an unacceptable increase in convergence time, as long as not all the parameters are unconstrained.

Assessing tension between uncorrelated datasets is relatively straightforward, but correlated datasets pose an implementation challenge.
We demonstrate how to perform such an analysis using ACT DR6 and \texttt{NPIPE} lensing in the \cobaya{} framework, where their correlations are known.

We use \pc{} to generate the posterior samples, and the tension analyses were performed using \ac{} \citep{polychord1, polychord2, anesthetic}.
\ac{} was also used to produce the posterior corner plots.

\section{Methods}

\subsection{Tools}
This work uses \pc{} \texttt{v1.20.2} to explore the parameter spaces and produce posterior samples.
The sampling and modelling framework \cobaya{} \citep{cobayaascl, Torrado2021} provides the interface between the likelihoods, sampler and the Boltzmann code \camb{} \texttt{v1.4.2.1} \citep{Lewis:1999bs, Lewis:2002ah, Howlett:2012mh}.
We use a modified version of \cobaya{} \texttt{v3.3.1} with improvements to the interface with \pc{}\footnote{\url{https://github.com/handley-lab/cobaya}}.
All nested sampling runs were performed with 1,000 live points starting from 10,000 prior samples.
A development version of the recently released \ac{} \texttt{v2} was used to create the posterior plots and compute tension statistics and Kullback--Leibler (KL) divergences \citep{dkl}.

\subsection{Quantifying tension between uncorrelated datasets}

\begin{table*} \centering
  \fontsize{8.5}{10}\selectfont  % Custom font size: 8.5pt text, 10pt line spacing
  \newcommand{\gr}{\cellcolor{gray!25}}  % shorthand for gray cell
  \makebox[\linewidth][c]{%  % Center table on full page width, allow extension beyond margins
    \begin{tabular}{|c|ccccccc|}
      \hline
      Data & Prior & $H_0 (\unit{\km \per \s \per \mega \parsec}) $ & $S_8 = \sigma_8 \sqrt{\Omega_\mathrm m / 0.3}$ & $\log{R}$ & $d$ & $\log{S}$ & $p$-value \\
      \hline
      ACT Vs Planck & uniform &&& $7.331 \pm 0.282$ & $3.299 \pm 0.434$ & $1.555 \pm 0.097$ & $97.7 \pm 2.6 \%$ \\
      \gr SPT-3G Vs Planck & \gr uniform &\gr&\gr& \gr $8.389 \pm 0.271$ & \gr $3.610 \pm 0.438$ & \gr $0.627 \pm 0.101$ & \gr $60.9 \pm 3.5 \%$ \\
      \hline
      & informative &&& $1.963 \pm 0.112$ & $1.030 \pm 0.100$ & $0.041 \pm 0.034$ & $34.0 \pm 2.4 \%$ \\
      \multirowcell{-2}{ACT Vs BAO} & \gr BBN &\gr&\gr& \gr $1.729 \pm 0.110$ & \gr $0.874 \pm 0.112$ & \gr $-0.025 \pm 0.037$ & \gr $29.6 \pm 2.8 \%$ \\
      \hline
      & informative &&& $1.914 \pm 0.114$ & $1.119 \pm 0.101$ & $-0.040 \pm 0.034$ & $30.8 \pm 2.2 \%$ \\
      \multirowcell{-2}{NPIPE Vs BAO} & \gr BBN &\gr&\gr& \gr $1.435 \pm 0.115$ & \gr $1.052 \pm 0.110$ & \gr $-0.260 \pm 0.038$ & \gr $22.2 \pm 2.0 \%$ \\
      \hline
      & informative &&& $1.594 \pm 0.101$ & $1.213 \pm 0.103$ & $-0.319 \pm 0.036$ & $22.0 \pm 1.7 \%$ \\
      \multirowcell{-2}{SPT-3G Vs BAO} & \gr BBN &\gr&\gr& \gr $1.612 \pm 0.103$ & \gr $0.434 \pm 0.116$ & \gr $0.249 \pm 0.038$ & \gr $91.7 \pm 16.3 \%$ \\
      \hline
      & informative &&& $6.633 \pm 0.118$ & $2.202 \pm 0.107$ & $0.779 \pm 0.036$ & $76.7 \pm 1.8 \%$ \\
      \multirowcell{-2}{ACT Vs NPIPE} & \gr BBN & \gr\multirow{-10}{*}[2pt]{\includegraphics[height=116pt]{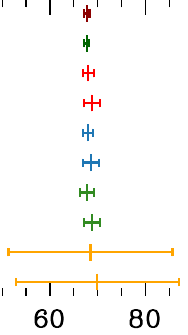}} & \gr\multirow{-10}{*}[2pt]{\includegraphics[height=116pt]{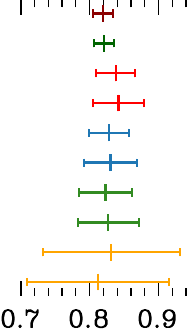}} & \gr $6.460 \pm 0.120$ & \gr $1.717 \pm 0.103$ & \gr $0.955 \pm 0.036$ & \gr $100.0 \pm 0.3 \%$* \\
      \hline

    \end{tabular}
    }% End makebox
    \vspace{6pt}
    \caption{
      $H_0$, $S_8$, Bayes' factor, Bayesian model dimensionality, suspiciousness and respective $p$-values for ACT lensing data combined with Planck anisotropies, BAO, DES-Y1 and Planck \texttt{NPIPE} lensing.
      ``Informative'' and ``uniform'' refer to the priors described in Table~\ref{tab:priors}.
      Only the uniform priors are used with Planck anisotropy measurements.
      All the $p$-values are well above 0.05, therefore none of the datasets are in tension with one another.
      Values of $S_8$ and $H_0$ shown are those found using the combined datasets.\\
      Notice that using uninformative priors has made only a very slight difference to the $S_8$ measurement by ACT + BAO.\\
      $^*$ Comparing ACT and \texttt{NPIPE} lensing, we find them in strong agreement, but following the $\log S$ prescription we find $p=100\%$.
      This is because the lower bound of the $p$-value integral in \ref{eq:p}, $d-2\log S$, is negative, and a $\chi^2$ distribution is normalised from zero to infinity.
      This may be interpreted in two ways: very strong agreement between the two lensing measurements, but also the Gaussian approximation used in the suspiciousness statistic breaking in this case where \lcdm{} is not well constrained by CMB lensing alone.
      The Bayes factor involves no such approximations, and $\log R>6$ for both informative and uniform priors reassures us that ACT and \texttt{NPIPE} lensing are not in tension.
    }
    \label{tab:sus}
\end{table*}

Cosmological models may be compared via their Bayes factor, calculated using the same dataset.
However, here we wish to compare two datasets to assess whether they are in tension.
This is usually achieved through the evidence ratio:
\begin{equation}
  R = \frac{Z_{AB}}{Z_A Z_B} \text,
  \label{eq:bayes}
\end{equation}
where $Z_A$ and $Z_B$ are the Bayesian evidences calculated using datasets $A$ and $B$ respectively, and $Z_{AB}$ is calculated using both datasets jointly.
$\log R > 0$ indicates concordance (the data are better explained by one shared universe than two separate ones), while $\log R < 0$ indicates tension.
As shown in Appendix~\ref{apx:langlelogSrangle}, the expectation value of $\log R$ under concordance is the mutual information between the datasets, which is positive-definite and prior-dependent.
This means that, besides comparison with the Jeffreys' scale \citep{jeffreys1939theory}, it is difficult to define a threshold for good agreement.
\citet{bevinscalibration} offers a recipe for calibrating $\log R$ using Neural Ratio Estimation; challenges with this approach will be addressed in Section~\ref{sec3:results}.

Suspiciousness aims to remove the prior dependence of $R$.
Coined in \citet{suspiciousness}, $S$ approximately removes the prior dependence from $R$ by dividing it by the ratio of the KL divergences between the posterior and the prior of each dataset.
After taking logs, suspiciousness can then be written as the difference of the expectation values of the log-likelihoods:
\begin{equation}
  \log{S} = \langle \log{\mathcal L_{AB}} \rangle_{\mathcal P_{AB}} - \langle \log{\mathcal L_{A}} \rangle_{\mathcal P_{A}} - \langle \log{\mathcal L_{B}} \rangle_{\mathcal P_{B}}\text,
\end{equation}
where $\langle \log{\mathcal L} \rangle_\mathcal P$ denotes the average value of (log) likelihood $\mathcal L$ over the posterior distribution $\mathcal P$.
Also in Appendix~\ref{apx:langlelogSrangle}, we show that the expectation value of $\log S$ under concordance vanishes, which makes the sign of $\log S$ a more calibrated indicator of tension than $\log R$.

In the case that the likelihood is exactly Gaussian, $d-2\log{S}$ has a $\chi_d^2$ distribution, where $d$ is the Bayesian model dimensionality of the shared parameters, defined in \citet{pablo}:
\begin{equation}
    \begin{aligned}
        \frac{\tilde{d}_D}{2} &= \langle(\log\mathcal L_{D})^2\rangle_{\mathcal P_D} - \langle\log\mathcal L_{D}\rangle_{\mathcal P_D}^2 \text,\\
        d &= \tilde d_A + \tilde d_B - \tilde d_{AB}\text,
    \end{aligned}
\end{equation}
where $\tilde d_D$ is the Bayesian model dimensionality of dataset $D$.
This is used to calculate the \textit{tension probability}, $p$, that discordance between the datasets is by chance:
\begin{equation}
  p = \int_{d-2\log{S}}^{\infty} \chi_d^2(x)\,\mathrm dx = \int_{d-2\log{S}}^{\infty} \frac{x^{d/2-1}e^{-x/2}}{2^{d/2}\mathrm\Gamma(d/2)}\,\mathrm dx \text.
  \label{eq:p}
\end{equation}
If $p\lesssim 0.05$ (corresponding to two Gaussian standard deviations) then the datasets are in moderate tension, while $p\lesssim 0.003$ corresponds to there being strong tension.

\subsection{Quantifying tension between correlated datasets}
\label{sec:correlated}
If two datasets have significant correlation, we must include this when comparing them, as outlined in \citet{correlated}.
Here, we demonstrate how this prescription is applied to the specific case of CMB lensing.
Equation~\ref{eq:bayes} is really making the following comparison:
\begin{equation}
  R = \frac{Z(\text{datasets } A \text{ and } B \text{ fit one universe together})}{Z(\text{datasets } A \text{ and } B \text{ fit one universe each})} = \frac{Z(H_0)}{Z(H_1)} \text,
\end{equation}
where $H_0$ is the null hypothesis that the two datasets are both measurements of the same universe (not to be confused with the Hubble constant in other sections); $H_1$ is the alternative hypothesis that they are each a measurement of a separate universe with different cosmological parameters.
The corresponding suspiciousness then becomes:
\begin{equation}
  \log S = \langle\log\mathcal L_{H_0}\rangle_{\mathcal P_{H_0}} - \langle\log\mathcal L_{H_1}\rangle_{\mathcal P_{H_1}}\text.
\end{equation}

The ACT DR6 lensing likelihood is Gaussian in the data, so the likelihood corresponding to $H_1$ is:
\begin{align}
  \log\mathcal L_{H_1} &= {\log\mathcal L_{H_1}}_\text{max}\label{eq:h1} \\\notag
  &- \frac 1 2
  \begin{bmatrix}
    C_\ell(\theta_A) - D_A \\ C_\ell(\theta_B) - D_B
  \end{bmatrix}^\intercal
  \begin{bmatrix}
    \mathrm\Sigma_A & \mathrm\Sigma_X \\ \mathrm\Sigma_X^\intercal & \mathrm\Sigma_B 
  \end{bmatrix}^{-1}
  \begin{bmatrix}
    C_\ell(\theta_A) - D_A \\ C_\ell(\theta_B) - D_B
  \end{bmatrix}
  \text.
\end{align}
$\theta_A$ and $\theta_B$ are the parameters corresponding to the two universes, which \cobaya{} provides to \camb{} to calculate the CMB power spectrum $C_\ell(\theta)$.
$D_A$ and $D_B$ are the power spectra measured by ACT and \texttt{NPIPE} respectively, $\mathrm\Sigma_A$ and $\mathrm\Sigma_B$ are the correlations within each dataset and $\mathrm\Sigma_X$ are the cross-correlations between the two.
The evidence for the alternative hypothesis is calculated by integrating over the doubled parameter space:
\begin{equation}
  Z(H_1) = \int\mathcal L_{H_1}(\theta_A, \theta_B) \pi(\theta_A)\pi(\theta_B)\,\mathrm d\theta_A\,\mathrm d\theta_B\text.
\end{equation}
One can see that $Z(H_1) = Z_AZ_B$ in the case the two datasets are uncorrelated, where the cross-correlations $\mathrm\Sigma_X=0$, and we recover the prescription in the previous section.
The likelihood for $H_0$ with a single set of parameters $\theta$ can then be obtained by fixing $\theta_A = \theta_B$:
\begin{equation}
  \mathcal L_{H_0}(\theta) = \mathcal L_{H_1}(\theta_A = \theta, \theta_B = \theta)\text,
\end{equation}
so the evidence for the null hypothesis is:
\begin{equation}
  Z(H_0) = \int\mathcal L_{H_0}(\theta)\pi(\theta)\,\mathrm d\theta\text.
\end{equation}

Out of the box, the ACT \cobaya{} likelihood only requests a single set of $C_\ell$s, corresponding to $\mathcal L_{H_0}$.
To compute $H_1$, we modified the likelihood to request two sets of $C_\ell$s using each half of the doubled parameter space.
More detail and the source code is also included in Appendix~\ref{apx:correlations} and the cookbook \citep{zenodo}.

\subsection{Priors}

\begin{figure*}
  \centering
  \includegraphics[width=0.49\textwidth]{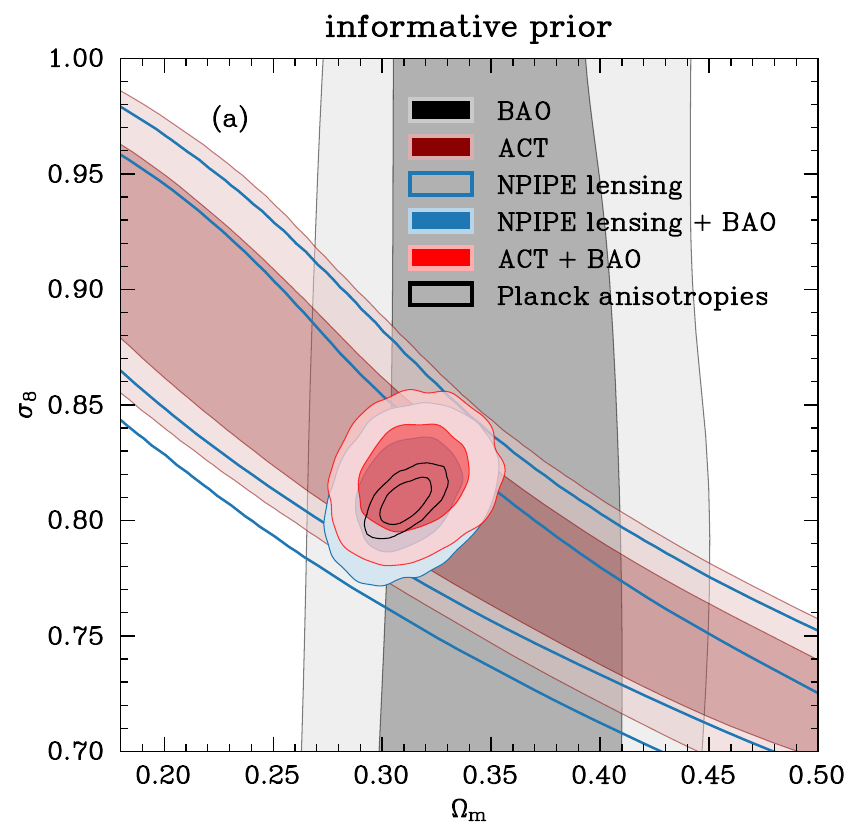}
  \includegraphics[width=0.49\textwidth]{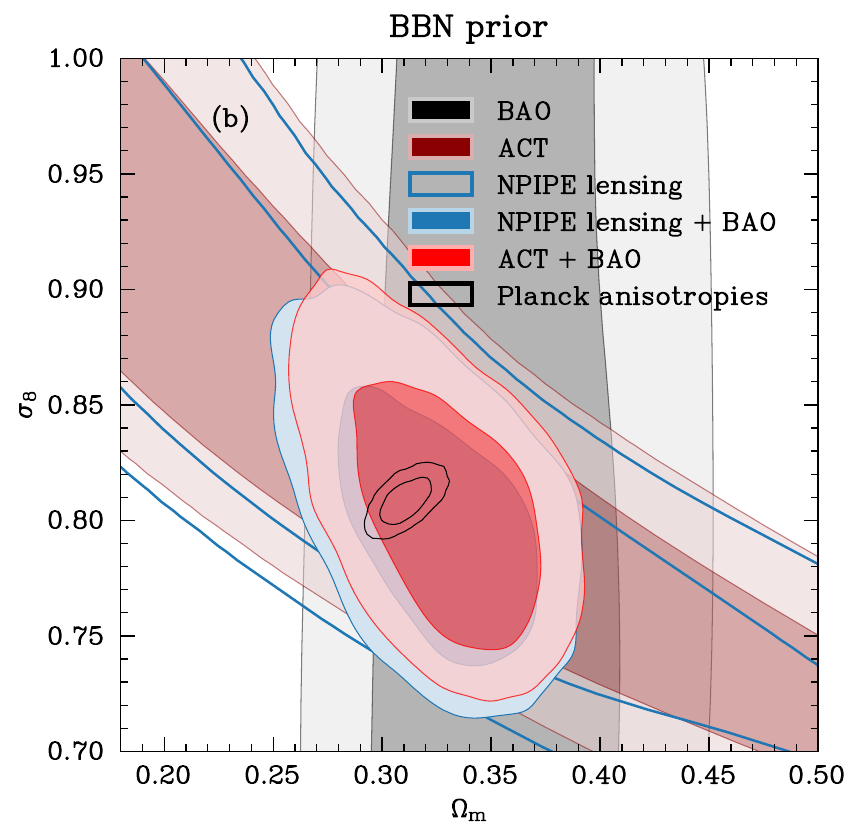}
  \caption{
    \textit{(a) left:} The amplitude of matter fluctuations $\sigma_8$ measured by ACT or \texttt{NPIPE} lensing, BAO, Planck lensing, and Planck CMB anisotropies.
    These were obtained using \pc{} and the informative priors used in \protect\citet{actdr6cosmo} on $\Omega_\textrm bh^2$, $n_\textrm s$, and fixed $\tau$.
    This is a repeat of Figure~6(a) from that paper.
    Lensing is degenerate in $H_0$, $\sigma_8$ and $\Omega_\textrm m$, which is broken by BAO measurements.
    The $\sigma_8$ value obtained by combining ACT with BAO agrees with \texttt{NPIPE} lensing + BAO, as well as Planck anisotropy measurements.
    However, it can be seen that the $1\sigma$ contours of the combined datasets are far tighter than the intersection of those of the separate datasets.
    It is not a problem in itself that the marginal joint posterior in the $\Omega_\mathrm m$ -- $\sigma_8$ plane is small when the individual marginals are large, since likelihood combination and marginalisation do not commute.
    However, it suggests correlations with other parameters, so the effect of the informative prior on $n_\mathrm s$ warranted further investigation.
    \textit{(b) Right:} Repeat of \textit{(a)}, using uniform priors on $n_\mathrm s$, retaining BBN information and fixed $\tau$.
    Combining the CMB lensing measurements with BAO results in $1\sigma$ and $2\sigma$ contours that better fill those of the separate datasets.\\
    The Planck anisotropy contours (with uniform CMB priors) are included on both figures for contrast, these include high-$\ell$ $TTTEEE$, low-$\ell$ $TT$ and low-$\ell$ $EE$ likelihoods.
  }
  \label{fig:actfig6}
\end{figure*}

\begin{figure*}
  \centering
  \includegraphics[width=0.49\textwidth]{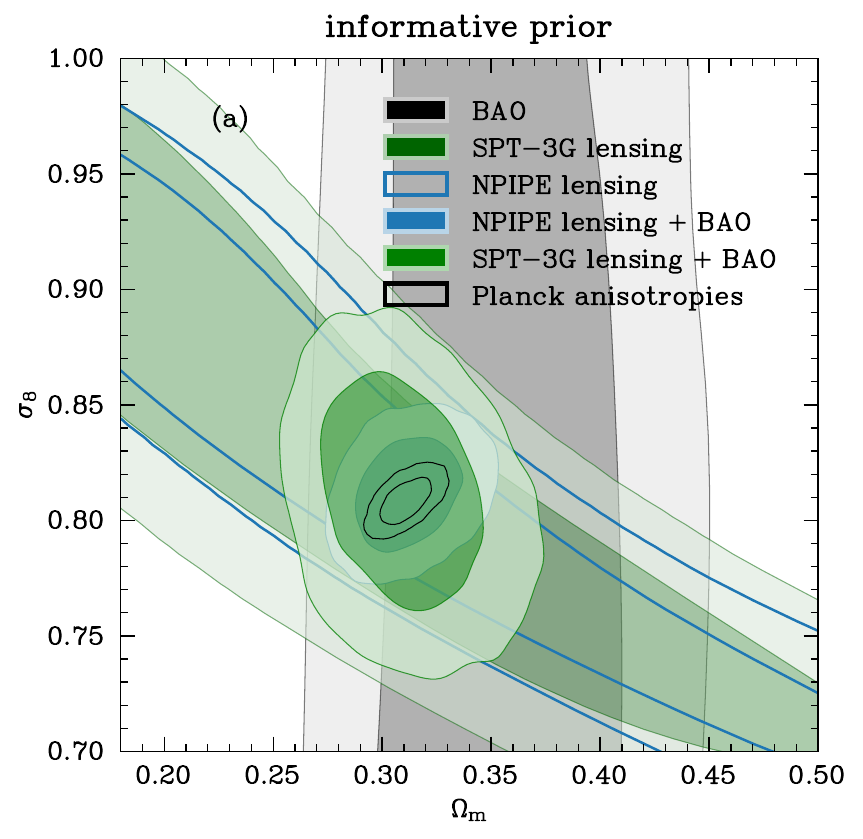}
  \includegraphics[width=0.49\textwidth]{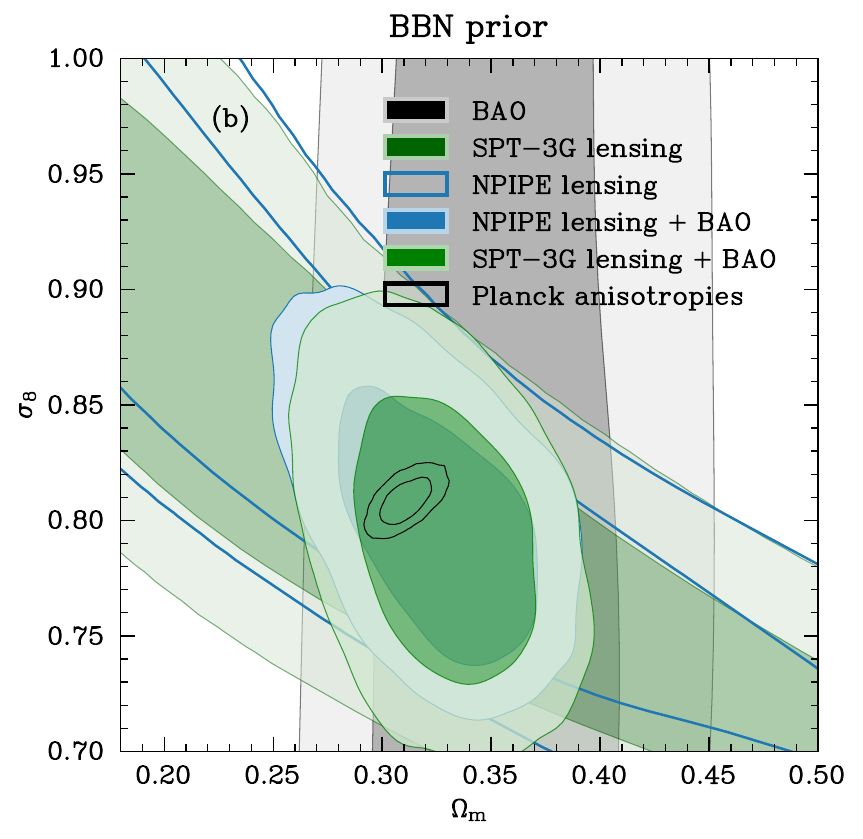}
  \caption{
    \textit{(a) left:} Repeat of Figure~\ref{fig:actfig6} (a), with SPT-3G instead of ACT lensing and retaining the informative prior.
    The constraints from SPT combined with BAO appear significantly wider than \texttt{NPIPE} lensing.
    \textit{(b) Right:} Similar repeat of Figure~\ref{fig:actfig6} \textit{(b)}, using uniform priors on $n_\mathrm s$ and the BBN prior on $\Omega_\mathrm bh^2$.
    With this prior, both lensing datasets produce similar constraints when combined with BAO.
  }
  \label{fig:sptfig6}
\end{figure*}

CMB lensing alone does not measure the baryonic density $\Omega_\mathrm bh^2$, the scalar spectral index of scalar fluctuations $n_\mathrm s$, or optical depth to reionisation $\tau$.
It is usual for CMB lensing measurements to be combined with big bang nucleosynthesis (BBN), which is achieved with a Gaussian prior on $\Omega_\mathrm b h^2$ \citep{bbn}.
Since $Z = \int\mathcal L\pi\,\mathrm d\theta$, multiplicative terms can be exchanged between the likelihood and prior without changing the theoretical evidence.
However, moving more information into the prior will generally improve sampling efficiency, for example, \pc{} uses the inverse cumulative distribution of the prior to transform directly from the unit hypercube to the prior.
Since this is straightforward for a Gaussian prior, it is sensible to incorporate BBN data in this way.

That leaves the spectral tilt.
\citet{planck15xv} use an informative prior on $n_\mathrm s$, motivated by the degeneracy between the power spectrum parameters $A_\mathrm s$ and $n_\mathrm s$.
This prior has been used in subsequent lensing analyses \citep{planck18viii, spt18, spt, act2seasons, actdelensed, actdr6cosmo} to provide equal footing for comparisons between the experiments.
However, we find that relaxing the $n_\mathrm s$ prior to a uniform prior typical of CMB primary analysis makes a significant difference to constraints in the $\Omega_\mathrm m$ -- $\sigma_8$ plane.
Appendix~\ref{apx:Asns} examines this further.

Finally, the optical depth does not affect the lensing power spectrum, so lensing analyses fix $\tau$ to a single value.
Introducing an additional unused parameter with a separable prior should not affect the evidence:
\begin{equation}
  Z = \int\mathcal L(\theta) \pi(\theta)\pi(\tau)\,\mathrm d\theta\,\mathrm d\tau = \int\mathcal L(\theta)\pi(\theta)\,\mathrm d\theta\int\pi(\tau)\,\mathrm d\tau\text.
\end{equation}
If the prior $\pi(\tau)$ is normalised, the right hand integral evaluates to 1, so it is unnecessary to spend computing effort to sample $\tau$.

\begin{table}
  \centering
  \rowcolors{2}{}{gray!25}
  \begin{tabular}{|l|c|c|c|}
    \hline
    & \multicolumn{3}{c|}{Prior} \\\cline{2-4}
    \multirow{-2}{*}{Parameter} & informative $n_\mathrm s$ & uniform $n_\mathrm s$ & CMB primary \\
    \hline
    $n_\mathrm s$ & $\mathcal N(0.96, 0.02^2)$ & \multicolumn{2}{c|}{$[0.8, 1.2]$} \\
    $\Omega_\mathrm b h^2$ & \multicolumn{2}{c|}{$\mathcal N(0.02233, 0.00036^2)$} & $[0.005, 0.1]$ \\
    $\tau$ & \multicolumn{2}{c|}{$0.055$} & $[0.01, 0.8]$ \\
    $\ln{10^{10} A_\mathrm s}$ & \multicolumn{3}{c|}{$[1.61, 4.0]$} \\
    $\Omega_\mathrm c h^2$ & \multicolumn{3}{c|}{$[0.005, 0.99]$} \\
    $100 \theta_\mathrm{MC}$ & \multicolumn{3}{c|}{$[0.5, 10]$} \\
    $H_0$* & \multicolumn{3}{c|}{$[40, 100]\unit{\km \per \s \per \mega \parsec}$} \\
    \hline
  \end{tabular}
  \caption{
    \lcdm{} parameter priors used in this work.
    Fixed values are indicated by a single number, uniform priors are denoted by brackets, and Gaussian priors as $\mathcal N(\mu, \sigma^2)$.
    The two sets of priors match those used in the ACT DR6 lensing analysis, the informative lensing prior used in lieu of any constraining power of $\Omega_\mathrm bh^2$, $n_\mathrm s$ and $\tau$ from CMB lensing and BAO \protect\citep{actdr6cosmo}.
    These closely follow previous analyses by Planck and SPT.
    The $\Omega_\mathrm bh^2$ lensing prior represents the result from BBN measurements \protect\citep{bbn}.
    The CMB prior is typical of priors used in tandem with CMB primary measurements.
    The third prior we consider in this work is the \textbf{BBN} prior, which combines the lensing prior with the uniform $n_\mathrm s$ CMB prior.\\
    * The $H_0$ prior is not sampled directly, but values outside this range are rejected by \camb{}.
  }
  \label{tab:priors}
\end{table}

\subsection{Datasets}
\subsubsection{ACT lensing}
ACT DR6 lensing results were published with a corresponding \cobaya{} likelihood.
This likelihood has four variants: ACT-only or ACT+Planck, each using either the baseline or extended multipole range.
We restrict this analysis to the baseline multipole range.
There is also an option \texttt{lens\_only} which must be set to false when combining with any primary CMB measurement.

ACT recommend a minimum set of \camb{} settings in their README\footnote{\url{https://github.com/ACTCollaboration/act_dr6_lenslike}}, these are the settings used here.
These can be found in Appendix~\ref{apx:camb}.

\subsubsection{\texttt{NPIPE} Planck lensing}
To assess the validity of combining ACT and Planck lensing measurements, we also require a separate Planck lensing likelihood.
Here we use the \texttt{NPIPE} Planck DR4 likelihoods \citep{NPIPE2, NPIPE1}.
Similarly to the ACT likelihoods, there are versions un-marginalised and marginalised over CMB measurements, which should be used with and without a separate CMB measurement respectively.

Since ACT and \texttt{NPIPE} lensing are measurements of the same sky, they have significant correlation, which means the likelihood corresponding to using both datasets is not the product of the likelihoods for each dataset individually.
This is addressed in the ACT likelihood, which we modified to take two sets of cosmological parameters, one for ACT and one for \texttt{NPIPE}.
This process is described in more detail in Section~\ref{sec:correlated}.

\subsubsection{SPT-3G}
Another CMB lensing experiment is the South Pole Telescope \citep{spt}.
Unfortunately, there is not a likelihood available which combines the SPT-3G lensing measurements with ACT or Planck taking into account correlations, so it is not possible to quantify the (unlikely) possibility of tension between them.

\subsubsection{Planck anisotropies}
There are a selection of Planck likelihoods for both low-$\ell$ (2<$\ell$<29) and high-$\ell$.
Since we are using the \texttt{NPIPE} Planck lensing, it seems appropriate to venture beyond the \texttt{plik} likelihood, which served as the baseline high-$\ell$ pipeline for the Planck 2018 legacy release \citep{planck18v}.
For low-$\ell$, we use the \texttt{clik} COMMANDER $TT$ and \texttt{SRoll2} $EE$ likelihoods, and for high-$\ell$ the \texttt{NPIPE} CamSpec combined $TTTEEE$ likelihood \citep{plancklowlTT, sroll2, camspecTTTEEE}.

CamSpec uses the Planck PR4 (\texttt{NPIPE}) maps released in 2020.
The \texttt{NPIPE} maps correspond to approximately $10\%$ tighter parameter constraints compared to the 2018 maps.
\texttt{SRoll2} also improves upon the 2018 low-$\ell$ $EE$ maps by improving the corrections for the nonlinear response of the analogue-to-digital converters of the Planck High Frequency Instrument, reducing the variance by a factor of two for $\ell < 6$, with improvements up to multipoles of 100.
The \texttt{clik} $TT$ likelihood uses the COMMANDER maps.

\subsubsection{BAO}
To break the degeneracy between $H_0$, $\sigma_8$ and $\Omega_\mathrm m$, CMB lensing measurements are combined with data from the 6df and SDSS surveys, specifically 6dFGS, SDSS DR7 Main Galaxy Sample (MGS), BOSS DR12 luminous red galaxies (LRGs) and eBOSS DR16 LRGs \citep{Beutler:2012px, Ross:2014qpa, Alam:2020sor}.
To be consistent with previous analyses we only include BAO information, which requires making a copy of the data files and covariance matrices from the native \cobaya{} eBOSS DR16 likelihood, removing the \texttt{fsigma8} elements from each, and providing these to a generic \cobaya{} BAO likelihood.
The data files and a copy of the \texttt{yaml} file are included in the cookbook \citep{zenodo}.

\section{Results}\label{sec3:results}

\begin{figure*}
  \centering
  \includegraphics[width=0.9\textwidth]{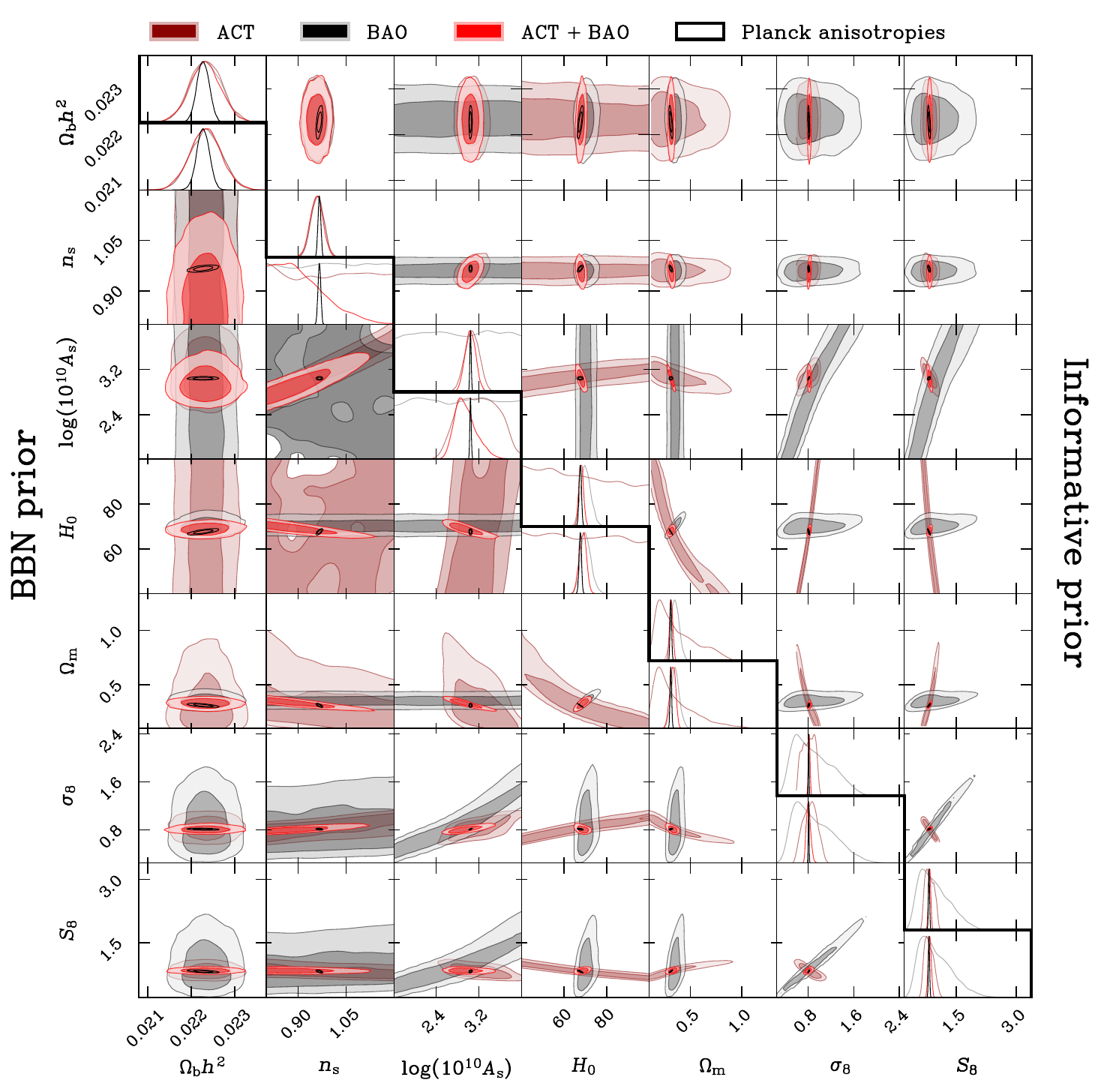}
  \caption{
    Corner plot comparing cosmological parameter constraints from ACT, BAO, and ACT + BAO, with informative and uniform priors on $n_\text s$.
    The plots with the informative prior are above the bold diagonal line, those below the diagonal use the uniform $n_\mathrm s$ prior.
    Constraints from Planck anisotropies are also shown for contrast, these use the uniform prior.
    Since $\tau$ is fixed, these panels are omitted from the informative prior.\\
    Note that $S_8$ is constrained well without the need for informative priors, also that the Planck constraints on $n_\text s$ are in fact tighter than the informative prior.
  }
  \label{fig:actcorner}
\end{figure*}

\begin{table}
  \centering
  \rowcolors{2}{}{gray!25}
  \begin{tabular}{|l|ccc|}
    \hline
    & \multicolumn{3}{c|}{$\mathcal D_\mathrm{KL}(\mathcal P || \pi)$} \\
    \cline{2-4}
    \multirow{-2}{*}{Prior} & ACT + BAO & \texttt{NPIPE} + BAO & SPT-3G + BAO \\
    \hline
    informative & $8.180 \pm 0.071$ & $8.281 \pm 0.074$ & $6.850 \pm 0.063$ \\
    BBN & $8.518 \pm 0.073$ & $8.609 \pm 0.074$ & $7.411 \pm 0.065$ \\
    \hline
    difference & $0.338 \pm 0.103$ & $0.328 \pm 0.106$ & $0.561 \pm 0.090$ \\
    \hline
  \end{tabular}
  \caption{
    Kullback--Leibler divergences between the posterior and prior $\mathcal D_\mathrm{KL}(\mathcal P || \pi)$ for each lensing dataset combined with BAO.
    The KL divergence quantifies the information gained by the posterior from the likelihood compared to the prior.
    Informative and BBN priors are used for each dataset, and the difference between the compressions for each prior is also shown.
    Values for ACT and \texttt{NPIPE} are similar, the latter containing marginally more information.
    SPT-3G contains less information than the other two, as the KL divergences are less than the other telescopes.
    The difference between using the informative versus BBN prior is also largest for SPT-3G: 0.56 nats less compression from prior to posterior when starting from the informative prior instead of the BBN prior, versus around 0.33 nats less compression with ACT or \texttt{NPIPE}.
  }
  \label{tab:dkl}
\end{table}

In Figure~\ref{fig:actfig6} (a), the combined lensing + BAO posterior contours are notably tighter in the $\Omega_\mathrm m$ -- $\sigma_8$ plane than the intersection of the individual dataset contours.
This is not problematic in itself, since combining likelihoods and projecting their posteriors do not commute.
It does, however, suggest that there are correlations with other parameters, motivating investigation of the effect of the informative prior on the spectral index $n_\mathrm s$.
Comparing Figures~\ref{fig:actfig6} (a) and (b), we see that a uniform prior on $n_\mathrm s$, typical of a CMB primary analysis, allows the ACT or \texttt{NPIPE} lensing + BAO $1\sigma$-contours to better fill the intersection between the individual contours of lensing and BAO in the $\Omega_\mathrm m$ -- $\sigma_8$ plane, compared to using the informative prior.
In Table~\ref{tab:sus}, we also see that the 1-D constraints on both $H_0$ and $S_8$ from each lensing + BAO have only changed slightly with the uniform $n_\mathrm s$ prior.

In Figure~\ref{fig:sptfig6} (a), the contours from SPT-3G combined with BAO are around twice as wide as those from the other two CMB lensing measurements.
Relaxing the prior to produce Figure~\ref{fig:sptfig6} (b) brings SPT much closer to the other CMB experiments.
We quantify this with Kullback--Leibler divergences, listed in Table~\ref{tab:dkl} \citep{dkl}.
ACT and \texttt{NPIPE} are very similar, with \texttt{NPIPE} approximately 0.1 nats ahead and gaining similar information from the informative prior.
SPT-3G contains over 1 fewer nats, and relaxing the $n_\mathrm s$ prior brings it closer to the other two.
This suggests that SPT-3G gains the most from using the informative prior.

Figure~\ref{fig:actcorner} demonstrates that, aside from the power spectrum parameters, the posterior for ACT + BAO is qualitatively unchanged by changing the prior on $n_\mathrm s$, in particular the $S_8$ measurement is virtually identical, so the uniform prior should not present significant extra difficulty for modern sampling tools.
Similar plots for \texttt{NPIPE} and SPT-3G can be found in Appendix~\ref{apx:corner}.

Bayes factors and tension probabilities $p$ were calculated for each of ACT, \texttt{NPIPE} and SPT-3G lensing versus BAO.
All $p$-values are comfortably greater than $5\%$, so there is no evidence for tension between them.
For both ACT and \texttt{NPIPE}, $p$ decreased when the $n_\mathrm s$ prior was relaxed, whereas SPT-3G has even stronger agreement with BAO using the relaxed prior.
ACT and SPT-3G were also compared with Planck anisotropies, both with good agreement.
Notably SPT has the greater Bayes factor but a lower $p$-value.
In every case $\log R>1$, supporting that these data are not in tension.

A methodology for calibrating $R$ is outlined in \citet{bevinscalibration} (\textsc{TensionNet}); this has not been carried out as it would require the use of forward models which can produce simulated pairs of observations from different experiments consistent with the same set of parameters.
This is challenging, particularly when CMB data are involved.
If one were to produce a simulated Planck-primary-ACT/SPT-lensing dataset pair, as required to train the Neural Ratio Estimator (NRE), it is not enough to just create a CMB map and a lensing map separately.
Because lensing is a physical remapping of the CMB photons, simulations must be statistically coupled by sharing a common underlying gravitational potential.
This ensures that the deflection field applied to the primary CMB is the same one recovered by the lensing reconstruction.
Otherwise, the simulations are of two different universes that happen to share cosmological parameters but have different realisations.
At low $\ell$, the observed CMB temperature and the lensing potential are also cross-correlated through the Integrated Sachs--Wolfe effect \citep{2006PhR...429....1L, NPIPE2}, as the photons traverse time-evolving gravitational potentials.
Independent simulations would fail to capture this cross-covariance.
If the simulated pairs do not share the same stochastic realisation, then this will bias the NRE training.
It will misinterpret the physical cross-correlation present in the real data but missing from the training set as tension (or possibly the absence of it), as it would learn incorrectly that ``concordance=zero cross-correlation''.

These difficulties are compounded by the need to faithfully model the complex instrumental effects of both the Planck satellite and the ACT telescope.
In particular, the ground-based ACT and SPT suffer noise from atmospheric and scan strategy effects, which makes it harder to forward-model than Planck's relatively well-characterised noise.
The simulated lensing map would need to be passed through the pipelines used by the experiments, further adding to the computational cost of producing the $\mathcal O(10^5)$ pairs of simulations recommended by \citet{bevinscalibration}.

Finally, we compared ACT lensing with \texttt{NPIPE} lensing, which required the methods described in Section~\ref{sec:correlated}.
For both priors we obtain Bayes factors $\log R > 6$, so we believe they also are not in tension.
However, with the uniform prior, we find $p=100\%$.
This is because the lower bound of the $p$-value integral, $d-2\log S$, is negative, and the $\chi^2$ distribution in Equation~\ref{eq:p} is of course normalised from zero to infinity.
This can be interpreted in two ways: lensing alone does not significantly constrain the \lcdm{} parameter space, so the Gaussian approximation used in the derivation of $p$ does not hold well, but also that the agreement between the two is so strong that it falls very deep in the tail of the $\chi^2$.
The Bayes factor makes no such approximations, and is the appropriate measure to use in this case.

A summary of parameter values and tension statistics are given in Table~\ref{tab:sus}; a more complete list of parameter values for both priors are included in Appendix~\ref{apx:params} for reference.
However, we recommend downloading and creating corner plots of the nested sampling chains to make proper comparisons, available on zenodo \citep{zenodo}.

\section{Conclusions}

In this work, nested sampling was used to produce posterior samples of \lcdm{} parameters using CMB lensing, BAO, BBN, and CMB primary measurements, using informative and uniform priors for the spectral index $n_\mathrm s$.
The suspiciously tight constraints on matter fluctuations resulting from the combination of ACT or \texttt{NPIPE} lensing with BAO become more reasonable with the uninformative prior.
SPT-3G is also more consistent with the results from ACT and \texttt{NPIPE} using the uniform prior.

Nested samples from both priors are presented on corner plots created using \texttt{anesthetic 2}, and were used to calculate the suspiciousness statistic between the three lensing datasets and BAO.
There is no substantial evidence for any tension between them.
We also find no tension between ACT or SPT-3G lensing with Planck anisotropies.

We have demonstrated that the informative prior on $n_\mathrm s$ makes a substantial difference to constraints in the $\Omega_\mathrm m$ -- $\sigma_8$ plane.
Therefore, we conclude that the informative priors used by the lensing community were not conservative, so we recommend that nested sampling with uninformative priors be used in future analyses of CMB lensing datasets.

As ACT lensing and \texttt{NPIPE} lensing measurements are correlated, quantifying their tension required modifying the ACT likelihood and \cobaya{} code to use two sets of cosmological parameters, one corresponding to each dataset.
We find that the suspiciousness $p$-value breaks down when comparing two datasets in strong agreement which together do not constrain the parameter space.
Therefore, we recommend that comparisons of such datasets are made using Bayes factors.
While tension was not expected, we have demonstrated how to perform such a correlated analysis between CMB datasets.

\section*{Acknowledgements}

This work was performed using the Cambridge Service for Data Driven Discovery (CSD3), part of which is operated by the University of Cambridge Research Computing on behalf of the STFC DiRAC HPC Facility (\url{www.dirac.ac.uk}).
The DiRAC component of CSD3 was funded by BEIS capital funding via STFC capital grants ST/P002307/1 and ST/R002452/1 and STFC operations grant ST/R00689X/1.
DiRAC is part of the National e-Infrastructure.

The tension calculations in this work made use of \textsc{NumPy} \citep{numpy}, \textsc{SciPy} \citep{scipy}, and \textsc{pandas} \citep{pandaszenodo, pandaspaper}.
The plots were rendered in \textsc{matplotlib} \citep{matplotlib}, using the \textsc{smplotlib} template created by \citet{smplotlib}.

%%%%%%%%%%%%%%%%%%%%%%%%%%%%%%%%%%%%%%%%%%%%%%%%%%
\section*{Data Availability}

All the nested sampling chains used in this analysis can be obtained from \citet{zenodo}.
This includes a \texttt{Jupyter} notebook demonstrating how we used \texttt{anesthetic} to create the figures and compute the tension statistics.



\bibliographystyle{mnras}
\bibliography{balancing_act} % if your bibtex file is called example.bib

%\begin{thebibliography}{99}
%\end{thebibliography}

%%%%%%%%%%%%%%%%%%%%%%%%%%%%%%%%%%%%%%%%%%%%%%%%%%

%%%%%%%%%%%%%%%%% APPENDICES %%%%%%%%%%%%%%%%%%%%%

\appendix

\section{$\langle\log R\rangle_{P(A,B)}$ and $\langle\log S\rangle_{P(A,B)}$}\label{apx:langlelogSrangle}

This appendix shows that, under the assumptions of uncorrelated likelihoods and concordance, the expectation value of $\log R$ over datasets $A$ and $B$ is equal to their mutual information, and that the expectation value of suspiciousness $\log S$ vanishes.
For easier arithmetic manipulation and pattern matching, the symbol $P$ is used for all probabilities, so the evidence $Z(A)=P(A)$ etc.
The tension ratio may be written:
\begin{equation}
    \begin{aligned}
        \log R &= \log\frac{P(A,B)}{P(A)P(B)}\text,\\
        \langle\log R\rangle_{P(A,B)} &= \int\log\frac{P(A,B)}{P(A)P(B)}P(A,B)\,\mathrm dA\,\mathrm dB\text.\\
    \end{aligned}
\end{equation}
This is recognised as the expression for the mutual information of $A$ and $B$ for continuous random data \citep{shannon, mutualinformation}.

The term which $\log R$ and $\log S$ differ by, $I$, is known as the ``information ratio'' in \citet{pablo}, though it has been discussed that ``interaction information'' may be a more accurate moniker.\footnote{\href{https://github.com/handley-lab/anesthetic/pull/333}{github.com/handley-lab/anesthetic/pull/333}, \href{https://github.com/handley-lab/anesthetic/pull/411}{github.com/handley-lab/anesthetic/pull/411}}
The $\log$ has also been dropped since information is a logarithmic quantity.
Its expectation value may be computed as follows:
\begin{equation}
    \begin{aligned}
        I &= \mathcal D_\mathrm{KL}(P(\theta|A)||P(\theta)) + D_\mathrm{KL}(P(\theta|B)||P(\theta))\\
        &- D_\mathrm{KL}(P(\theta|A,B)||P(\theta))\text,\\
        \langle I\rangle_{P(A,B)} &= \iint\left[\int\log\frac{P(\theta|A)}{P(\theta)}P(\theta|A)\,\mathrm d\theta \right.\\
        &\quad+\int\log\frac{P(\theta|B)}{P(\theta)}P(\theta|B)\,\mathrm d\theta \\
        &\quad\left.- \int\log\frac{P(\theta|A,B)}{P(\theta)}P(\theta|A,B)\,\mathrm d\theta\right]P(A,B)\,\mathrm dA\,\mathrm dB\text.
    \end{aligned}
\end{equation}
Use Bayes theorem to replace $P(A|\theta)/P(\theta)=P(\theta|A)/P(A)$ etc., and use the fact that the first two terms do not depend on $B$ and $A$ respectively, and combine the measures.
\begin{equation}
    \begin{aligned}
        &\langle I\rangle_{P(A,B)} = \iiint\left[\log\frac{P(A|\theta)}{P(A)} + \log\frac{P(B|\theta)}{P(B)}\right.\\
        &\qquad\qquad\qquad\qquad-\left.\log\frac{P(A,B|\theta)}{P(A,B)} \right]P(A,B, \theta)\,\mathrm d\theta\,\mathrm dA\,\mathrm dB\\
        &= \iiint\log\left[\frac{P(A|\theta)P(B|\theta)}{P(A,B|\theta)}\frac{P(A,B)}{P(A)P(B)}\right]P(A,B,\theta)\,\mathrm d\theta\,\mathrm dA\,\mathrm dB\text.
    \end{aligned}
\end{equation}
However, we are considering the uncorrelated likelihood case, so $P(A,B|\theta)=P(A|\theta)P(B|\theta)$, and the first fraction cancels.
This leaves no dependence on $\theta$, so:
\begin{equation}
    \begin{aligned}
        \langle I\rangle_{P(A,B)} &= \int\log\frac{P(A,B)}{P(A)P(B)}P(A,B)\,\mathrm dA\,\mathrm dB\\
        &= \langle\log R\rangle_{P(A,B)} \text,\\
        \therefore \langle\log S\rangle_{P(A,B)} &= \langle\log R\rangle_{P(A,B)} - \langle I\rangle_{P(A,B)} = 0 \text,
    \end{aligned}
\end{equation}
and thus, the expectation value of $\log S$ has vanished.

\section{Correlated datasets with \texttt{Cobaya}}
\label{apx:correlations}
As outlined in Section~\ref{sec:correlated}, in order to calculate the evidence for the alternative hypothesis of tension between ACT and \texttt{NPIPE} lensing, one needs to simultaneously sample over two sets of cosmological parameters used to calculate two sets of $C_\ell$s.
Adapting \cobaya{} to sample two sets of cosmological parameters was not straightforward, so the process is outlined here.

\cobaya{} uses dictionaries to record the current set of parameter values, therefore, two sets of cosmological parameters are most easily distinguished by having different names.
For simplicity, since ACT contains fewer characters than \texttt{NPIPE}, it was decided that the ACT portion should use the renamed quantities which were all prefixed ``\texttt{ACT}'', and \texttt{NPIPE} would use parameters with the usual labels.

\cobaya{} also makes use of caching partial results.
In particular, the transfer functions do not depend on the power spectrum parameterised by $A_\mathrm s$ or $n_\mathrm s$, so \cobaya{} caches the most recent transfer functions in a \texttt{CambTransfers} object.
The sampler considers these then as ``fast'' parameters as it is efficient to consider several different values of those while keeping $\Omega_\mathrm bh^2$, $\Omega_\mathrm ch^2$, $\theta_\mathrm{MC}$ and $\tau$ constant.
This performance benefit would be difficult to preserve if the same objects were responsible for two sets of cosmological parameters, so the most straightforward solution was to duplicate the class.

First, the python files which interface with \camb{} were copied.
\cobaya{} is designed to be modular so that it is straightforward to introduce new theories and likelihoods, however, the class \texttt{BoltzmannBase} takes care of most of the interfacing common to different Boltzmann codes, which the \camb{} interface inherits.
A copy of \texttt{BoltzmannBase} was made and renamed to \texttt{ACTBoltzmannBase}, which \texttt{ACTCAMB} inherits.
The prefix ``ACT'' was then meticulously added as a prefix to all parameters, getters and setters, and other API elements to ensure the two \camb{}s were operating independently.

The \texttt{act\_dr6\_lenslike} likelihood also had to be modified.
The native likelihood requires ``\texttt{Cl}''s; the modified likelihood also requires ``\texttt{ACTCl}''s from \cobaya{}, which are the $C_\ell$s calculated using the parameters prefixed ``\texttt{ACT}'' and calculated by \texttt{ACTCAMB}.
These are then passed to the generic likelihood function, which was also modified to position the two sets of $C_\ell$s to perform the multiplication in Equation~\ref{eq:h1}.

This pipeline was tested by setting the ``\texttt{ACT}''-prefixed parameters equal to the usual parameters in the \texttt{yaml} file to calculate the evidence of the null hypothesis, and asserting this gave identical results to the native pipeline.

\section{Effect of $A_\mathrm s$ and $n_\mathrm s$ on the lensing power spectrum}\label{apx:Asns}

\begin{figure*}
    \begin{center}
        \includegraphics[width=0.95\textwidth]{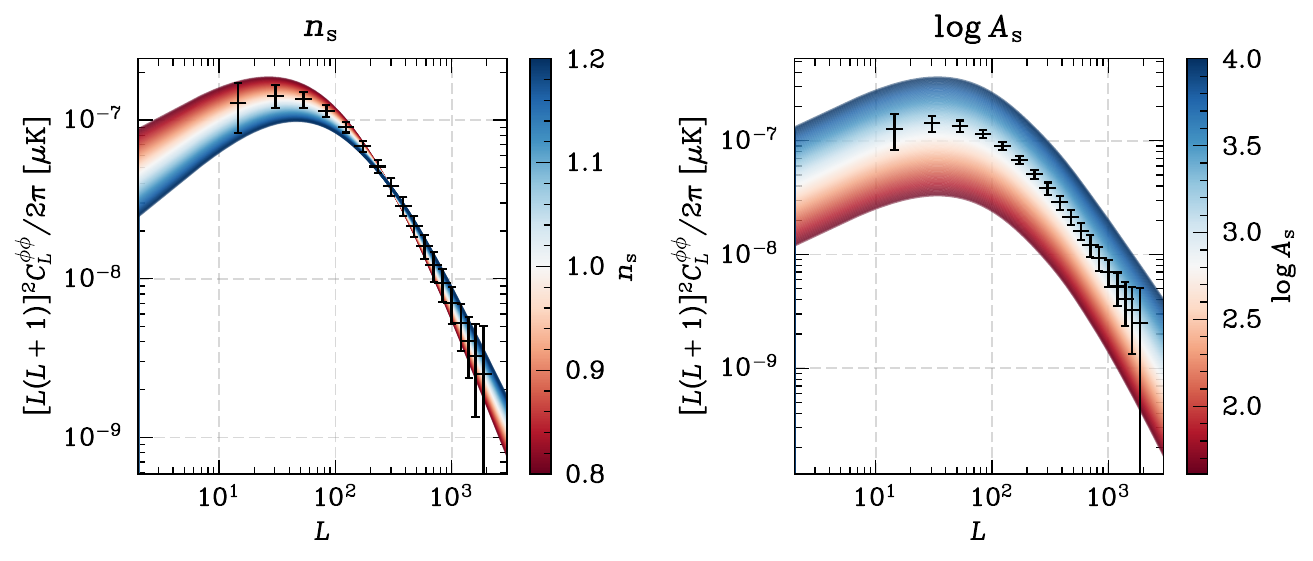}
    \end{center}
    \caption{
        CMB lensing power spectra, scanned over the priors for the power spectrum spectral tilt $n_\mathrm s$ and amplitude $A_\mathrm s$.
        As expected, the whole lensing spectrum shifts up in proportion to $A_\mathrm s$, but lensing power is shifted to large scales (low $\ell$) for lower $n_\mathrm s$, and vice versa for great $n_\mathrm s$.
        In each plot, all other parameters are fixed to their Planck 2018 values.
        The ACT DR6 lensing bandpowers and error bars are shown for reference \citep{actdr6cosmo}.
    }\label{fig:cmblensingcl}
\end{figure*}

\citet{actdr6cosmo} stated that the informative prior is necessary as the power spectrum parameters are degenerate with only a CMB lensing measurement.
Figure~\ref{fig:cmblensingcl} shows that scanning the uninformative priors of $\log A_\mathrm s$ and $n_\mathrm s$, while keeping the other parameters fixed at their Planck 2018 values \citep{planck2018VI}, have different effects on the lensing power spectrum $C_L^{\phi\phi}$.
The amplitude $A_\mathrm s$ simply shifts the whole spectrum up or down, whereas $n_\mathrm s$ tilts the spectrum, shifting power to large scales for lower $n_\mathrm s$, and vice versa for greater $n_\mathrm s$.
The ACT DR6 lensing bandpowers and error bars are shown for reference \citep{actdr6cosmo}, which, by inspection, appear to favour a slight red tilt.
The degeneracy between $A_\mathrm s$ and $n_\mathrm s$ is not exact, so there should certainly be no trouble for nested sampling to explore the posterior with uninformative priors.

\section{CAMB settings}\label{apx:camb}
ACT recommend a minimum set of \texttt{CAMB} settings in their README; these are the settings used here in the \texttt{yaml} format used by \texttt{Cobaya}.

\begin{minted}[fontsize=\small]{yaml}
theory:
  camb:
    stop_at_error: False
    extra_args:
      bbn_predictor: PArthENoPE_880.2_standard.dat
      halofit_version: mead2016
      lens_potential_accuracy: 4
      lmax: 4000
      lens_margin: 1250
      AccuracyBoost: 1
      lSampleBoost: 1
      lAccuracyBoost: 1
      nnu: 3.046
      num_massive_neutrinos: 1
      theta_H0_range:
      - 40
      - 100
\end{minted}

\section{\texttt{NPIPE} and SPT-3G corner plots}
\label{apx:corner}
\begin{figure*}
  \centering
  \includegraphics[width=0.9\textwidth]{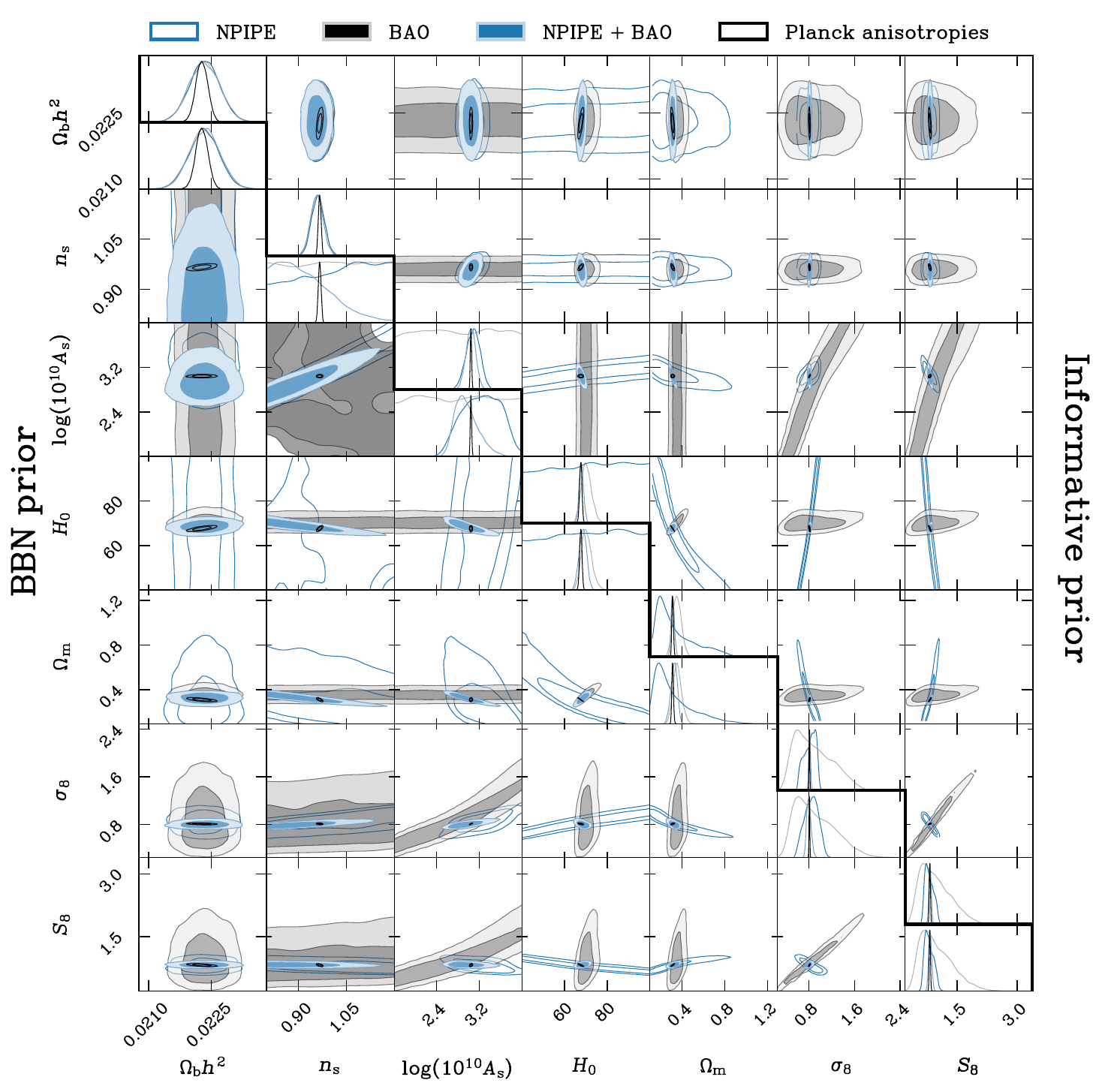}
  \caption{
    Corner plot comparing cosmological parameter constraints from \texttt{NPIPE}, BAO, and \texttt{NPIPE} + BAO, with informative and uniform priors on $n_\mathrm s$.
    The plots with the informative prior are above the bold diagonal line, those below the diagonal use the uniform $n_\mathrm s$ prior.
    The distributions are virtually identical to Figure~\ref{fig:actcorner}.
  }
\end{figure*}

\begin{figure*}
  \centering
  \includegraphics[width=0.9\textwidth]{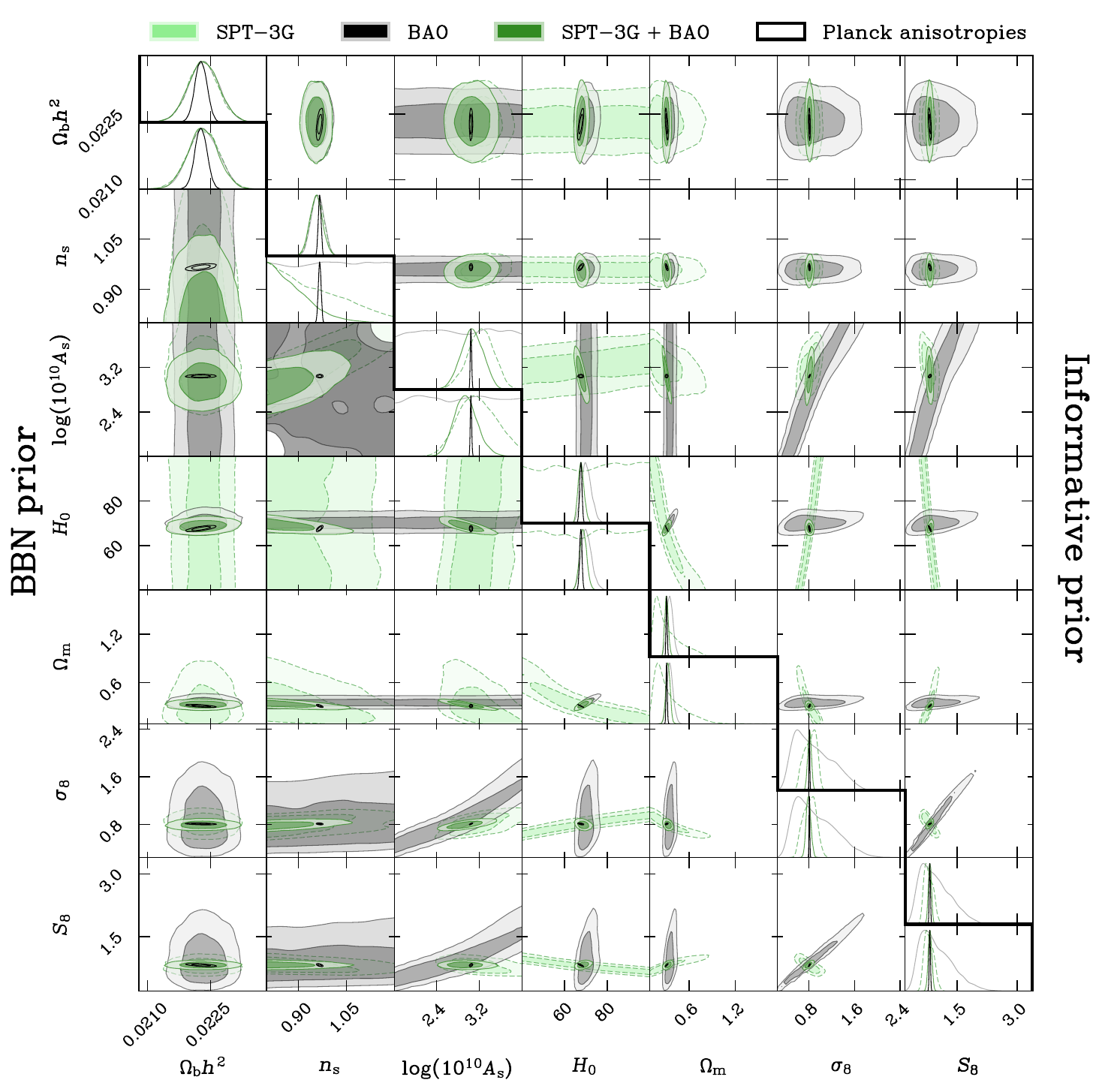}
  \caption{
    Corner plot comparing cosmological parameter constraints from SPT-3G, BAO, and SPT-3G + BAO, with informative and uniform priors on $n_\mathrm s$.
    The plots with the informative prior are above the bold diagonal line, those below the diagonal use the uniform $n_\mathrm s$ prior.
    Note particularly that the Planck contours involving $n_\mathrm s$ do not overlap the $1\sigma$ contours from SPT-3G + BAO.
  }
\end{figure*}

\section{Parameter constraints}\label{apx:params}

For reference, the following table lists the numeric values of a selection of cosmological parameters for every dataset and combination thereof used in this work.

\begin{table*}
  \centering
  \begin{tabular}{lcccc}
    & $\sigma_8$ & $S_8 = \sigma_8 \sqrt{\Omega_\mathrm m / 0.3}$ & $\Omega_\mathrm m$ & $H_0 (\unit{\km \per \s \per \mega \parsec})$ \\
    \hline
    \textbf{informative prior}\\
    ACT & $0.805 \pm 0.102$ & $0.840 \pm 0.103$ & $0.371 \pm 0.188$ & $67.8 \pm 17.3$ \\
    \texttt{NPIPE} & $0.816 \pm 0.098$ & $0.812 \pm 0.102$ & $0.336 \pm 0.176$ & $70.9 \pm 17.2$ \\
    SPT-3G & $0.823 \pm 0.104$ & $0.795 \pm 0.106$ & $0.318 \pm 0.169$ & $70.6 \pm 17.5$ \\
    ACT + \texttt{NPIPE} & $0.806 \pm 0.100$ & $0.832 \pm 0.100$ & $0.361 \pm 0.182$ & $68.5 \pm 17.2$ \\
    BAO & $0.889 \pm 0.323$ & $0.971 \pm 0.377$ & $0.354 \pm 0.040$ & $70.3 \pm 2.4$ \\
    ACT + BAO & $0.819 \pm 0.015$ & $0.838 \pm 0.028$ & $0.314 \pm 0.016$ & $68.0 \pm 1.1$ \\
    ACT + BAO & $0.819 \pm 0.015$ & $0.838 \pm 0.028$ & $0.314 \pm 0.016$ & $68.0 \pm 1.1$ \\
    \texttt{NPIPE} + BAO & $0.812 \pm 0.016$ & $0.828 \pm 0.029$ & $0.312 \pm 0.016$ & $67.9 \pm 1.1$ \\
    SPT-3G + BAO & $0.811 \pm 0.032$ & $0.823 \pm 0.038$ & $0.310 \pm 0.024$ & $67.8 \pm 1.4$ \\
    \hline
    \textbf{BBN prior}\\
    ACT & $0.833 \pm 0.119$ & $0.828 \pm 0.109$ & $0.343 \pm 0.194$ & $69.4 \pm 17.3$ \\
    \texttt{NPIPE} & $0.843 \pm 0.118$ & $0.806 \pm 0.106$ & $0.317 \pm 0.180$ & $70.8 \pm 17.1$ \\
    SPT-3G & $0.813 \pm 0.110$ & $0.804 \pm 0.111$ & $0.336 \pm 0.185$ & $70.2 \pm 17.5$ \\
    ACT + \texttt{NPIPE} & $0.843 \pm 0.118$ & $0.812 \pm 0.103$ & $0.324 \pm 0.184$ & $69.9 \pm 17.1$ \\
    BAO & $0.909 \pm 0.338$ & $0.995 \pm 0.397$ & $0.355 \pm 0.040$ & $70.4 \pm 2.4$ \\
    ACT + BAO & $0.807 \pm 0.037$ & $0.842 \pm 0.037$ & $0.328 \pm 0.028$ & $68.8 \pm 1.7$ \\
    ACT + BAO & $0.807 \pm 0.037$ & $0.842 \pm 0.037$ & $0.328 \pm 0.028$ & $68.8 \pm 1.7$ \\
    \texttt{NPIPE} + BAO & $0.801 \pm 0.038$ & $0.830 \pm 0.038$ & $0.324 \pm 0.029$ & $68.6 \pm 1.7$ \\
    SPT-3G + BAO & $0.793 \pm 0.040$ & $0.827 \pm 0.044$ & $0.327 \pm 0.027$ & $68.8 \pm 1.7$ \\
    \hline
    \textbf{uniform prior}\\
    Planck anisotropies & $0.809 \pm 0.007$ & $0.823 \pm 0.015$ & $0.311 \pm 0.008$ & $67.6 \pm 0.6$ \\
    ACT + Planck & $0.807 \pm 0.006$ & $0.820 \pm 0.015$ & $0.309 \pm 0.008$ & $67.7 \pm 0.6$ \\
    SPT-3G + Planck & $0.808 \pm 0.006$ & $0.821 \pm 0.014$ & $0.310 \pm 0.007$ & $67.7 \pm 0.6$ \\
  \end{tabular}
  \caption{
    $\sigma_8$, $S_8$, $\Omega_\mathrm m$ and $H_0$ values from the datasets and priors used in this work.
  }
\end{table*}

%%%%%%%%%%%%%%%%%%%%%%%%%%%%%%%%%%%%%%%%%%%%%%%%%%

% Don't change these lines
\bsp	% typesetting comment
\label{lastpage}
\end{document}